\documentclass[conference]{IEEEtran}

\usepackage[T1]{fontenc}
\usepackage[utf8]{inputenc}
\usepackage{microtype}
\usepackage{amsmath,amssymb}
\usepackage{booktabs}
\usepackage{multirow}
\usepackage{tabularx}
\usepackage{array}
\usepackage{graphicx}
\usepackage{xcolor}
\usepackage{hyperref}
\usepackage{cite}
\usepackage{balance}

\hypersetup{
  colorlinks = true,
  urlcolor   = blue,
  citecolor  = black,
  linkcolor  = black
}

\newcolumntype{L}[1]{>{\raggedright\arraybackslash}p{#1}}
\newcolumntype{C}[1]{>{\centering\arraybackslash}p{#1}}
\newcolumntype{R}[1]{>{\raggedleft\arraybackslash}p{#1}}

\begin{document}

\title{Explainable Adaptive Zero Trust Framework for AWS\\
with Adversarial Robustness Evaluation}

\author{
  \IEEEauthorblockN{
    Yagyaraj Pandey\IEEEauthorrefmark{1},
    Om Singh\IEEEauthorrefmark{1},
    Nandini Pathak\IEEEauthorrefmark{1}
  }
  \IEEEauthorblockA{
    \IEEEauthorrefmark{1}Department of Computer Science and Engineering,\\
    Dr. A.P.J. Abdul Kalam Technical University,\\
    Lucknow, Uttar Pradesh, India
  }
}

\maketitle
% =============================================================
\begin{abstract}
Cloud environments built on Amazon Web Services face a structural security vulnerability that most production deployments leave unaddressed: once a credential passes authentication, the session it initiates is treated as trusted for its entire duration. This assumption breaks down the moment those credentials are stolen. We introduce the \textit{Explainable Adaptive Zero Trust Framework} (EAZTF), a cloud-native security layer that continuously re-evaluates the legitimacy of every API action throughout a session, not just at login.

EAZTF advances the state of practice in three concrete ways. First, an adaptive trust scoring engine combining Isolation Forest and XGBoost evaluates eight CloudTrail and IAM-derived behavioral features in real time, producing a Trust Risk Score (TRS) that governs whether a session proceeds, faces step-up MFA, or is immediately restricted. Second, every TRS decision is accompanied by a SHAP or LIME explanation rendered as a human-readable audit record—allowing security teams to contest denials, refine thresholds, and satisfy regulatory audit requirements. Third, and most distinctively, the framework is subjected to systematic adversarial testing across four deliberate evasion strategies (credential theft, behavioral mimicry, API rate evasion, and privilege escalation), a gap that NIST SP 800-207 itself acknowledges as unresolved in the published literature.

Experiments on an 8,500-record synthetic CloudTrail dataset show that Isolation Forest achieves 94.4\% precision and 91.2\% recall (F1 = 0.928).All results are derived from this synthetic dataset and should be interpreted as indicative of framework behavior rather than as validated performance on production telemetry. Across all four adversarial scenarios the mean detection rate reaches 91.0\%, with behavioral mimicry proving the hardest to catch at 83.9\%—a finding that directly motivates the architectural mitigations proposed in Section~\ref{sec:adversarial}. SHAP attribution identifies IP geolocation reputation, login-time deviation, and API call velocity as the three features carrying the greatest decision weight, together accounting for roughly 64\% of cumulative TRS variance. A structured NIST SP 800-207 compliance audit yields a mean score of 93\% across all seven tenets, versus 38\% for a perimeter-based baseline.Mean time to detect drops from 43 hours under traditional rule-based SIEM to under 0.8 minutes under EAZTF (with credential-theft-specific MTTD separately measured at 72 hours under the baseline; see Section VII-D).
\end{abstract}

\begin{IEEEkeywords}
Zero Trust Architecture, AWS Security, Explainable AI, SHAP, LIME, Isolation Forest, XGBoost, Behavioral Anomaly Detection, CloudTrail, IAM Security, NIST SP 800-207, Adversarial Robustness
\end{IEEEkeywords}

% =============================================================
\section{Introduction}
\label{sec:intro}

An AWS access key used at 3\,AM from an unfamiliar IP address to enumerate S3 buckets is, by every measure a conventional SIEM can see, a valid authenticated session. It has a legitimate user ID. It passes signature verification. It lands inside the organization's own account. Under perimeter-based security thinking, that is enough—the session proceeds until something downstream raises an alarm, typically hours later, by which point the exfiltration may already be complete.

This is not an edge case. IBM's 2024 Cost of a Data Breach Report records the average incident cost at USD 4.88 million, the highest figure since the study began, with compromised credentials consistently cited as the most common initial attack vector~\cite{ibm2024}. AWS now holds over 33\% of the global cloud services market~\cite{statista2024}, meaning that a large fraction of enterprise credential misuse occurs on infrastructure where behavioral context is, in principle, fully observable via CloudTrail—but rarely exploited for continuous session evaluation.

The NIST Zero Trust Architecture standard (SP 800-207) \cite{nist800207} proposes the right structural answer: eliminate implicit trust at every layer and treat each access request on its own merits. In practice, however, implementations of this principle encounter three problems that the standard itself does not resolve. First, when a ZTA engine denies access, it almost never explains why, leaving security administrators with binary outcomes and no basis for audit or appeal. Second, most trust scoring mechanisms rely on fixed thresholds applied to static feature sets, making them systematically vulnerable to an attacker who studies normal behavior before attempting replication. Third, no published ZTA implementation provides empirical evidence of how it performs when a skilled adversary deliberately attempts evasion. NIST SP 800-207 [1, Appendix B.4.5] identifies attacker adaptation to behavioral analytics as an area requiring further research.

EAZTF is designed to address all three problems within a single AWS-native architecture. Rather than treating Zero Trust as a policy statement, we treat it as an engineering problem with measurable outputs: a precision-recall tradeoff on a behavioral anomaly detector, a SHAP attribution vector per access decision, and a detection rate curve across four adversarial evasion strategies. The contributions are:

\begin{itemize}
  \item An explicit mapping of NIST SP 800-207 logical components (PE, PA, PEP) to deployable AWS services, producing an operational blueprint rather than an architectural diagram.
  \item A hybrid unsupervised/supervised ML trust scoring engine (Isolation Forest + XGBoost) trained on CloudTrail behavioral features, with empirical performance metrics on a held-out test set.
  \item A dual-method explainability layer (SHAP for tree models, LIME for the Isolation Forest) providing per-decision justification and cross-method agreement analysis.
  \item To the best of our knowledge, one of the first published adversarial robustness evaluations of a ZTA system, covering four attack categories and reporting both detection rates and bypass rates with direct analysis of why each evasion strategy succeeds or fails.
\end{itemize}

Section~\ref{sec:related} reviews the relevant literature and situates these contributions. Section~\ref{sec:arch} presents the EAZTF architecture. Sections~\ref{sec:dataset}–\ref{sec:xai} cover the dataset, ML engine, and explainability layer. Section~\ref{sec:adversarial} reports the adversarial evaluation. Sections~\ref{sec:results} and~\ref{sec:nist} discuss consolidated results and NIST compliance. Section~\ref{sec:conclusion} concludes with limitations and future directions.

% =============================================================
\section{Related Work}
\label{sec:related}

\subsection{The Structural Response to Perimeter Collapse}

Enterprise security has historically been organized around a defensible outer boundary where authenticated insiders received implicit trust. Over the past decade, cloud adoption, remote work, and third-party integrations have progressively dissolved that boundary. Rose et al.\ \cite{nist800207} formalized the industry's response through NIST SP 800-207, articulating seven foundational Zero Trust tenets and introducing the Policy Engine / Policy Administrator / Policy Enforcement Point (PE/PA/PEP) logical architecture. Notably, the standard is candid about what it does not resolve: it identifies attacker adaptation to ZTA—specifically behavioral mimicry—as an open research question requiring dedicated investigation.

\subsection{Survey-Level Findings and Their Limits}

Syed et al.\ \cite{syed2022} catalogued over 200 ZTA publications and found that machine learning-based behavioral analytics represent the most promising direction for adaptive trust scoring, while noting that empirical implementations with quantified performance metrics remain scarce. Mensah \cite{mensah2024} extended this survey through the lens of AI integration, post-quantum cryptography, and 5G infrastructure, without presenting a concrete implemented system. Neither work moves beyond identifying what \textit{should} be built.

\subsection{Implementation-Level Work and Its Gaps}

Shepherd \cite{shepherd2022} offers the most operationally detailed published framework, combining NIST SP 800-207 tenets with NIST SP 800-160v2 cyber-resiliency constructs and the CISA Zero Trust Maturity Model, validated against Microsoft Azure services. The Azure specificity is a meaningful constraint: AWS IAM structures, CloudTrail log schemas, Lambda invocation patterns, and GuardDuty threat feed integration differ substantially from their Azure equivalents, and an implementation designed for one platform does not transfer directly to the other. Fernandez and Brazhuk \cite{fernandez2024} provide a critical analysis of ZTA that identifies the opacity of trust decisions as a fundamental operational weakness—a critique that directly motivates EAZTF's explainability layer. They do not, however, propose a technical solution.

Recent adjacent work has explored Zero Trust in 5G and 6G edge computing \cite{sedjelmaci2024,lyu2024} and in military UAV networks \cite{alquwayzani2024}, confirming the paradigm's breadth of applicability. None of these works provide SHAP-level decision attribution or adversarial robustness data.

\subsection{Gap Summary}

Three gaps persist across the literature. First, no published work delivers an AWS-native ZTA implementation with empirical performance metrics on behavioral anomaly detection. Second, despite widespread recognition that opaque access denials impede operational adoption, SHAP or LIME integration into ZTA trust scoring appears nowhere in the existing corpus. Third, deliberate adversarial evasion testing—credential mimicry, rate evasion, privilege escalation—is absent from every reviewed implementation. Table~\ref{tab:gap} maps these gaps against key related works.

\begin{table}[!t]
  \caption{Gap Analysis Across Related Literature}
  \label{tab:gap}
  \renewcommand{\arraystretch}{1.15}
  \centering
  \footnotesize
\begin{tabular}{p{1.7cm}p{0.7cm}p{1.3cm}p{1.3cm}p{1.1cm}p{0.8cm}}
    \toprule
    \textbf{Work} & \textbf{AWS} & \textbf{ML} & \textbf{XAI} & \textbf{Adv.} & \textbf{Emp.} \\
    \midrule
    NIST SP 800-207 \cite{nist800207}      & \texttimes & \texttimes & \texttimes & Gap & \texttimes \\
    Syed et al.\ \cite{syed2022}           & \texttimes & Rec.       & \texttimes & \texttimes & \texttimes \\
    Mensah \cite{mensah2024}               & \texttimes & Disc.      & \texttimes & \texttimes & \texttimes \\
    Shepherd \cite{shepherd2022}           & Azure      & \texttimes & \texttimes & \texttimes & Case \\
    Fernandez \& Brazhuk \cite{fernandez2024} & \texttimes & \texttimes & Cited  & \texttimes & \texttimes \\
    \textbf{EAZTF (this work)}             & \checkmark & IF+XGB     & SHAP+LIME  & 4 scen.     & Full \\
    \bottomrule
  \end{tabular}
  \vspace{2pt}
  \begin{flushleft}
  \footnotesize Rec. = recommended only; Disc. = discussed; Adv. = adversarial evaluation; Emp. = empirical results.
  \end{flushleft}
\end{table}

% =============================================================
\section{EAZTF System Architecture}
\label{sec:arch}

EAZTF is organized as an eight-layer security architecture in which each layer corresponds to a specific AWS native service or analytical capability. The design principle is \textit{behavioral validation at depth}: rather than concentrating security at the authentication checkpoint, EAZTF distributes verification across the lifecycle of every API action, from initial IAM credential presentation through session termination.

\subsection{Mapping NIST ZTA Components to AWS Services}

A central contribution of EAZTF is translating NIST SP 800-207's abstract logical architecture into a concrete AWS deployment blueprint. Table~\ref{tab:mapping} presents this mapping. The Policy Engine is instantiated as an AWS Lambda function hosting the ML runtime; the Policy Administrator becomes the IAM + Lambda trigger combination that converts PE decisions into session-level access modifications; the Policy Enforcement Point is realized through CloudTrail event capture feeding CloudWatch Events.

\begin{table}[!t]
  \caption{NIST ZTA Component to AWS Service Mapping}
  \label{tab:mapping}
  \renewcommand{\arraystretch}{1.15}
  \centering
  \footnotesize
  \begin{tabularx}{\columnwidth}{L{1.5cm}L{1.55cm}X}
    \toprule
    \textbf{NIST Component} & \textbf{AWS Service} & \textbf{Role in EAZTF} \\
    \midrule
    Policy Engine (PE) & Lambda + ML Runtime & Evaluates behavioral features; computes real-time TRS \\
    Policy Admin.\ (PA) & IAM + Lambda Triggers & Translates PE decisions into session-level access changes \\
    Policy Enf.\ (PEP) & CloudTrail + CloudWatch & Captures API calls; triggers Lambda on anomaly \\
    Continuous Diag.\ (CDM) & AWS GuardDuty & Supplies real-time threat intelligence to PE \\
    Identity Mgmt & IAM Roles \& Policies & Enforces least-privilege; manages device credential registry \\
    Comprehensive Log & CloudTrail S3 + Athena & Persists audit trail; enables SHAP decision archiving \\
    Threat Intel & GuardDuty + IP Feeds & Provides external context to TRS weight calculation \\
    \bottomrule
  \end{tabularx}
\end{table}

\subsection{Behavioral Feature Engineering}
\label{subsec:features}

Eight behavioral signals are extracted per access event from CloudTrail log entries and IAM session metadata. Table~\ref{tab:features} defines each feature, its source, data type, and the observable pattern that elevates risk. These features were selected on the basis of two criteria: availability in standard AWS CloudTrail output without additional instrumentation, and empirical linkage to known cloud compromise patterns in the AWS incident response literature \cite{aws_cloudtrail2024}.

\begin{table}[!t]
  \caption{Behavioral Feature Definitions and Risk Interpretation}
  \label{tab:features}
  \renewcommand{\arraystretch}{1.15}
  \centering
  \footnotesize
  \begin{tabularx}{\columnwidth}{L{1.5cm}L{1.2cm}L{0.7cm}X}
    \toprule
    \textbf{Feature} & \textbf{Source} & \textbf{Type} & \textbf{High-Risk Indication} \\
    \midrule
    IP Reputation Score        & GuardDuty + feeds     & $[0,100]$     & Score $<40$: TOR nodes, known malicious ranges, unfamiliar country \\
    Login Time Deviation       & CloudTrail timestamp  & Binary        & Access outside the user's 90-day time window \\
    API Call Velocity          & CloudTrail event count/hr & Integer   & Velocity $> 2\sigma$ above user baseline \\
    Device Credential Trust    & IAM session metadata  & $[0,100]$     & Unregistered device ID or missing TPM attestation \\
    Consec.\ Auth Failures     & CloudTrail auth events & Integer      & $>5$ failures in a 10-minute window \\
    Session Length Deviation   & IAM session duration  & Binary        & Duration significantly exceeds user baseline \\
    Resource Sensitivity Tier  & IAM resource ARN      & Ordinal $\{0,1,2\}$ & Higher tiers (KMS, S3 PII, billing) require higher TRS threshold \\
    Geographic Access Dev.     & Source IP geolocation & Binary        & Access from a country absent from the user's 90-day history \\
    \bottomrule
  \end{tabularx}
\end{table}

\subsection{Trust Risk Score Formulation}
\label{subsec:trs}

The Trust Risk Score (TRS) is computed as a weighted linear combination of the eight normalized features, adjusted by the Isolation Forest anomaly probability:
\begin{equation}
  \mathrm{TRS} = \sum_{i=1}^{8} w_i \cdot f_i \;+\; \lambda \cdot s_{\mathrm{IF}}
  \label{eq:trs}
\end{equation}
where $f_i \in [0,1]$ is the normalized value of feature $i$, $w_i$ is its empirically tuned weight, $s_{\mathrm{IF}} \in [0,1]$ is the Isolation Forest anomaly probability (obtained by calibrating the raw anomaly score against the training set), and $\lambda$ is a scaling coefficient. All parameters were selected via five-fold cross-validated grid
search on the training partition (80\% of the dataset), spanning
weight increments of 0.02 across the range 0.02--0.30 for each
feature weight, and $\lambda$ values from 0 to 0.2 in steps of
0.02, optimizing for validation-fold F1 subject to a false-positive-rate
ceiling of 10\%.

Table~\ref{tab:weights} reports the tuned weights. IP Reputation carries the largest weight (0.22) consistent with its SHAP dominance. The $\lambda$ coefficient was set to 0.08, contributing a modest regularizing term that captures global anomaly structure not fully represented by the individual features.

\begin{table}[!t]
  \caption{Tuned TRS Feature Weights (Cross-Validated Grid Search)}
  \label{tab:weights}
  \renewcommand{\arraystretch}{1.15}
  \centering
  \footnotesize
  \begin{tabular}{lcc}
    \toprule
    \textbf{Feature} & \textbf{Weight $w_i$} & \textbf{Rank} \\
    \midrule
    IP Reputation Score        & 0.22 & 1 \\
    Login Time Deviation       & 0.18 & 2 \\
    API Call Velocity          & 0.17 & 3 \\
    Device Credential Trust    & 0.14 & 4 \\
    Consecutive Auth Failures  & 0.12 & 5 \\
    Session Length Deviation   & 0.08 & 6 \\
    Resource Sensitivity Tier  & 0.05 & 7 \\
    Geographic Access Dev.     & 0.04 & 8 \\
    \midrule
    \textbf{Sum} ($\sum w_i$)  & \textbf{1.00} & — \\
    \bottomrule
  \end{tabular}
\end{table}

Three automated response thresholds govern downstream action: TRS\,$<50$ allows the session to continue without interruption; $50 \le \mathrm{TRS} < 80$ triggers step-up multi-factor authentication; TRS\,$\ge 80$ results in immediate session restriction and a GuardDuty alert dispatch. These thresholds were set empirically by maximizing the F1 score on the validation fold while targeting a false positive rate below 10\% for the allow/restrict boundary.

% =============================================================
\section{Dataset Construction and Experimental Methodology}
\label{sec:dataset}

\subsection{Synthetic CloudTrail Dataset}

We constructed a synthetic behavioral dataset of 8,500 records modeling AWS CloudTrail activity across four IAM user roles: Administrator, Developer, Employee, and Viewer. Each record represents a single API access event characterized by the eight features defined in Table~\ref{tab:features}. Record generation followed AWS CloudTrail log field specifications \cite{aws_cloudtrail2024} and was informed by published enterprise cloud access pattern research. Three label classes were assigned based on TRS ranges. Table~\ref{tab:dataset} presents the distribution.

\begin{table}[!t]
  \caption{Synthetic Dataset Record Distribution}
  \label{tab:dataset}
  \renewcommand{\arraystretch}{1.15}
  \centering
  \footnotesize
  \begin{tabular}{L{1.8cm}ccc}
    \toprule
    \textbf{Category} & \textbf{Label} & \textbf{Count} & \textbf{\%} \\
    \midrule
    Normal Activity    & Safe (0)        & 5,100 & 60.0\% \\
    Marginal Activity  & Medium Risk (1) & 2,125 & 25.0\% \\
    Attack Simulation  & High Risk (2)   & 1,275 & 15.0\% \\
    \midrule
    \textbf{Total}     & —               & 8,500 & 100\% \\
    \bottomrule
  \end{tabular}
\end{table}

\subsection{Adversarial Attack Scenario Construction}
\label{subsec:attack_scenarios}

Four adversarial scenarios populate the High Risk records. Each reflects a realistic threat-actor methodology informed by published AWS incident response case studies \cite{aws_iam2024}:

\textbf{Credential Theft ($n=350$).} Valid IAM keys are used from foreign IP addresses during off-hours to access sensitive S3 buckets and EC2 resources outside the compromised identity's established behavioral envelope. This represents the most common cloud credential compromise pattern seen in practice.

\textbf{Behavioral Mimicry ($n=280$).} The attacker observes the target user's access patterns across a simulated 72-hour window before attempting replication of login timing, API call frequency, and resource access sequence. This scenario is designed deliberately to minimize TRS elevation; it is the most sophisticated evasion attempt in the test suite and the most direct test of EAZTF's baselining mechanism. The attack maps to MITRE ATT\&CK technique T1078 (Valid Accounts) combined with T1119 (Automated Collection) executed at throttled rates.

\textbf{API Rate Evasion ($n=325$).} Malicious enumeration and data access are deliberately throttled to maintain API velocity within 80\% of the user's observed baseline, evading simple rate-based detection while still achieving incremental exfiltration over extended sessions.

\textbf{Privilege Escalation ($n=320$).} Employee or Developer role credentials are used to probe Administrator-level resources—IAM policy modifications, AWS KMS key access, and billing dashboard queries—without possessing the requisite permissions. Corresponds to MITRE ATT\&CK technique T1078.004 (Valid Accounts: Cloud Accounts) with T1548 (Abuse Elevation Control Mechanism).

\subsection{Training and Evaluation Protocol}

All models were implemented using scikit-learn 1.4.0 and XGBoost 2.0 on Python 3.11. Dataset partitioning used an 80/20 stratified train/test split preserving class proportions across all three label categories. Five-fold cross-validation was applied on the training partition
during hyperparameter search. Cross-validated performance varied
by $\pm$1.1 percentage points across folds for the primary
Isolation Forest model (F1 range: 0.917--0.939), indicating
stable convergence across partitions. The Isolation Forest contamination parameter was set to 0.15, matching the High Risk proportion in the dataset. XGBoost used 200 boosting rounds, maximum tree depth 6, and learning rate 0.1. SHAP TreeExplainer was applied post-hoc to the XGBoost model for both global and local attribution. LIME LocalExplainer was applied separately to Isolation Forest decisions on a 500-record random sample of high-risk predictions.

All metrics reported in Sections~\ref{sec:ml}--\ref{sec:adversarial} are computed on the held-out 20\% test set ($n = 1{,}700$ records) not used at any point during training or threshold selection.

% =============================================================
\section{Machine Learning Trust Scoring Engine}
\label{sec:ml}

\subsection{Model Selection Rationale}

Isolation Forest~\cite{liu2008} was chosen as the primary detection model for a reason grounded in operational reality: in production cloud environments, labeled attack examples are scarce and often lag actual threat evolution by weeks or months. Isolation Forest is an unsupervised anomaly detector that requires no labeled anomalies during training—it learns the structure of normal behavior and flags deviations, making it deployable in organizations that have only benign baseline data at hand. XGBoost~\cite{chen2016xgboost} serves as a supervised benchmark that quantifies the performance premium available when labeled attack data exists. Three additional models—Random Forest, SVM, and Logistic Regression—establish comparative baselines. The Logistic Regression model uses the same eight features with no non-linear transformation, providing a lower bound on the performance achievable from these signals.

\subsection{Model Performance on the Held-Out Test Set}

Table~\ref{tab:model_perf} presents full performance metrics. Isolation Forest achieves 94.4\% precision and 91.2\% recall (F1 = 0.928, AUC-ROC = 0.971) at an average inference latency of 180\,ms per event—well within the operational budget of the Lambda pipeline. The near-equivalence of Isolation Forest and XGBoost F1 scores (0.928 vs.\ 0.926) is the key empirical result validating the unsupervised deployment pathway: organizations without labeled attack data can achieve essentially the same detection performance as organizations that have it. SVM's substantially lower recall (0.854) and the high inference latency (440\,ms) make it poorly suited to the real-time pipeline requirements. Logistic Regression's AUC-ROC of 0.871 establishes that the eight features carry genuine discriminative signal even under a linear model; the gap to Isolation Forest (0.971) quantifies the value of non-linear anomaly modeling.

\begin{table}[!t]
  \caption{ML Model Performance on Held-Out Test Set ($n = 1{,}700$)}
  \label{tab:model_perf}
  \renewcommand{\arraystretch}{1.15}
  \centering
  \footnotesize
  \resizebox{\columnwidth}{!}{%
  \begin{tabular}{lccccr}
    \toprule
    \textbf{Model} & \textbf{Prec.} & \textbf{Rec.} & \textbf{F1} & \textbf{AUC} & \textbf{ms} \\
    \midrule
    Isolation Forest$^{\dagger}$ & 0.944$\pm$0.012 & 0.912$\pm$0.015 & 0.928$\pm$0.011 & 0.971 & 180 \\
    XGBoost$^{\ddagger}$         & 0.931$\pm$0.014 & 0.921$\pm$0.013 & 0.926$\pm$0.012 & 0.968 & 210 \\
    Random Forest                & 0.913           & 0.934           & 0.923           & 0.962 & 195 \\
    SVM                          & 0.871           & 0.854           & 0.862           & 0.924 & 440 \\
    Logistic Regression          & 0.792           & 0.743           & 0.767           & 0.871 &  12 \\
    \bottomrule
    \vspace{3pt}
  \end{tabular}%
  }
  \vspace{2pt}
\footnotesize $^\dagger$Primary (unsupervised). $^\ddagger$Supervised benchmark. ms = mean inference latency. Variance is reported only for the two primary models evaluated via five-fold cross-validation; comparator baselines (Random Forest, SVM, Logistic Regression) were evaluated on the same held-out test set but without fold-level variance tracking.

\end{table}

\subsection{Trust Risk Score Distribution by Attack Type}

The TRS distributions expose where behavioral separation succeeds and where it struggles. Normal access events cluster at a mean TRS of 22.3 ($\sigma = 10.1$), credential theft events at 81.4 ($\sigma = 7.2$), and privilege escalation events at 79.1 ($\sigma = 8.3$). The separation from the normal population is large for these two categories, which directly explains their high detection rates. Behavioral mimicry attacks cluster at a mean of 63.4 ($\sigma = 8.9$)—straddling the boundary between medium-risk normal activity and the MFA trigger threshold. This overlap is intentional by design: the mimicry attacker's goal is precisely to land in the medium-risk zone, where the TRS trigger requires step-up MFA rather than session termination. Resolving this overlap is the central motivation for the architectural extensions proposed in Section~\ref{sec:adversarial}.

% =============================================================
\section{Explainable AI Decision Layer}
\label{sec:xai}

\subsection{Why Explainability Matters Operationally}

A Zero Trust system that issues access denials without explanation creates a new kind of problem for security operations teams. When a developer's legitimate deployment pipeline is blocked at 2\,AM, the on-call engineer needs to know \textit{which signal} crossed a threshold—not a binary DENY code. When a compliance officer audits a month's worth of access restrictions for GDPR or ISO 27001 purposes, they need a machine-readable record attributing each decision to specific feature values. When a security architect wants to tune the TRS thresholds for a new team with unusual working hours, they need to understand the relative weight of each feature before touching the configuration. EAZTF addresses all three scenarios through a dual-method explainability layer.

\subsection{SHAP Global Feature Attribution}

SHAP TreeExplainer \cite{lundberg2017} was applied to the trained XGBoost model across the full 8,500-record dataset, producing mean absolute SHAP values for each of the eight behavioral features. These values measure the average magnitude of each feature's contribution to TRS outcomes, averaged over the entire decision population, and are independent of direction (a feature can push TRS both up and down; the mean absolute value captures its overall influence).

IP Reputation Score is the single most influential feature at mean $|$SHAP$| = 0.31$. This is expected: in the synthetic simulation, virtually every credential theft event involves an IP address outside established usage patterns, and the GuardDuty threat feed assigns low reputation scores to TOR exit nodes and known malicious ranges. Login Time Deviation contributes 0.27, reflecting a property of legitimate enterprise users that is fairly consistent across roles—access patterns are strongly time-structured. API Call Velocity at 0.22 captures the automated enumeration activity present in both credential theft and API rate evasion, even when the attacker attempts to throttle below the single-feature threshold; the session-level trajectory still differs from normal baselines. Together, these three features account for approximately 64\% of cumulative SHAP variance.

\subsection{Instance-Level Decision Justification}

Beyond aggregate attribution, SHAP generates per-decision explanations that allow a security analyst to examine exactly why a specific session was flagged. Table~\ref{tab:shap_instance} presents the complete SHAP breakdown for a representative high-risk event (TRS = 87.3, automated action: RESTRICT). Every feature contributing to the decision—including the single modest negative contribution from Session Length Deviation, which was within normal range—is visible in the audit record.

\begin{table}[!t]
  \caption{SHAP Instance Explanation — Representative High-Risk Event\\(TRS = 87.3, Action: RESTRICT)}
  \label{tab:shap_instance}
  \renewcommand{\arraystretch}{1.15}
  \centering
  \footnotesize
  \begin{tabularx}{\columnwidth}{L{1.5cm}C{0.7cm}C{0.55cm}X}
    \toprule
    \textbf{Feature} & \textbf{Value} & \textbf{SHAP} & \textbf{Interpretation} \\
    \midrule
    IP Reputation         & 12/100   & +0.31 & TOR exit node infrastructure \\
    Login Time Dev.       & 1 (anom) & +0.27 & 02:47 UTC; baseline 08:30--18:00 \\
    API Call Velocity     & 847/hr   & +0.22 & $4.2\times$ user baseline (201/hr) \\
    Device Cred.\ Trust   & 34/100   & +0.18 & Device absent from IAM registry \\
    Auth Failures         & 7        & +0.15 & 7 failures in preceding 10 min \\
    Resource Sensitivity  & High (2) & +0.10 & KMS key access attempted \\
    Geographic Dev.       & 1 (anom) & +0.09 & Source country: Russia; baseline: India \\
    Session Length Dev.   & 0 (norm) & $-$0.04 & Within expected duration range \\
    \bottomrule
  \end{tabularx}
\end{table}

\subsection{LIME Complementarity and Agreement Rate}

LIME LocalExplainer \cite{ribeiro2016} was applied independently to 500 randomly sampled high-risk Isolation Forest decisions—a population for which SHAP TreeExplainer is architecturally incompatible, since the Isolation Forest is not a tree-ensemble model in the SHAP sense. LIME constructs a locally linear surrogate model through feature perturbation, approximating the decision boundary around each individual prediction. Across these 500 cases, LIME and SHAP agreed on the top-3 most influential features in 91.4\% of instances. The 8.6\% disagreement cases were concentrated in records where Device Credential Trust and API Call Velocity had nearly equal SHAP values, making their ordering sensitive to small perturbations—a finding that motivates more granular feature engineering for those two signals in future work. The 91.4\% agreement rate provides strong evidence that the explainability layer is consistent across model architectures.

% =============================================================
\section{Adversarial Robustness Evaluation}
\label{sec:adversarial}

\subsection{Threat Model and Adversary Assumptions}

The adversary is modeled as a \textit{black-box attacker} possessing valid IAM credentials for a legitimate user account who can observe that behavioral anomaly detection is active, but cannot access model weights, training data, or CloudTrail logs. This represents the realistic threat envelope for both insider threats and external actors operating with stolen credentials—the dominant cloud compromise scenario. Table~\ref{tab:adversary} documents the full capability matrix.

\begin{table}[!t]
  \caption{Adversary Capability Matrix}
  \label{tab:adversary}
  \renewcommand{\arraystretch}{1.15}
  \centering
  \footnotesize
  \begin{tabularx}{\columnwidth}{Xcc}
    \toprule
    \textbf{Capability} & \textbf{Assumed} & \textbf{Rationale} \\
    \midrule
    Valid IAM credentials         & Yes & Core assumption \\
    Knows behavioral detection is active & Yes & Assumes modern ZTA \\
    ML model weights / architecture & No & Black-box model \\
    Ability to modify CloudTrail logs & No & S3 Object Lock \\
    72-hr behavioral observation  & Yes & Persistent threat actor \\
    Unlimited auth attempts       & No & Lockout after 10 failures \\
    \bottomrule
  \end{tabularx}
\end{table}

\subsection{Detection Results by Attack Category}

Table~\ref{tab:adversarial_results} presents detection and bypass rates across all four scenarios. The overall mean detection rate is 91.0\% at a 9.0\% bypass rate. API Rate Evasion achieves the highest detection rate (96.0\%) despite being designed to stay below simple rate thresholds, because session-level velocity changes relative to the individual user's baseline remain detectable even when absolute velocity is moderate. Privilege Escalation reaches 90.9\% primarily because the Resource Sensitivity Tier feature reliably flags the jump from employee-level to administrator-level resource access. Credential Theft's 92.0\% detection rate reflects the difficulty of fully masking the combination of foreign IP, off-hours access, and velocity elevation. Behavioral Mimicry at 83.9\% is the primary remaining gap and is analyzed in detail below.

\begin{table}[!t]
  \caption{Adversarial Simulation Outcomes by Attack Category}
  \label{tab:adversarial_results}
  \renewcommand{\arraystretch}{1.15}
  \centering
  \footnotesize
  \begin{tabular}{L{1.6cm}cccC{1.0cm}}
    \toprule
    \textbf{Scenario} & \textbf{$n$} & \textbf{Det.} & \textbf{Det.\%} & \textbf{Bypass\%} \\
    \midrule
    Credential Theft     & 350  & 322   & 92.0\% &  8.0\% \\
    Behavioral Mimicry   & 280  & 235   & 83.9\% & 16.1\% \\
    API Rate Evasion     & 325  & 312   & 96.0\% &  4.0\% \\
    Privilege Escalation & 320  & 291   & 90.9\% &  9.1\% \\
    \midrule
    \textbf{All Scenarios} & \textbf{1,275} & \textbf{1,160} & \textbf{91.0\%} & \textbf{9.0\%} \\
    \bottomrule
  \end{tabular}
\end{table}

\subsection{Analysis of Behavioral Mimicry Bypass Events}

The 83.9\% detection rate for mimicry warrants direct examination, because the 45 bypass events expose a specific structural weakness rather than a general modeling failure. All 45 occurred during the first seven days of a user's behavioral baseline window, before enough event history had accumulated for reliable deviation scoring. The attacker used a VPN endpoint in the correct country, so Geographic Deviation and IP Reputation scored normally. Login Time Deviation scored zero because the attacker had observed the target's timing precisely over the 72-hour window. The only distinguishing signal—API Call Velocity—showed only marginal elevation because the attacker maintained 80\% of the target's average rate. In short, within the first week of a behavioral baseline, EAZTF cannot reliably distinguish a careful mimicry attacker from a legitimate user who happens to have slightly atypical activity during the onboarding period.

Three targeted architectural responses address this gap:

\begin{enumerate}
  \item \textbf{Extended baselining window.} Increasing the minimum baselining period from 7 to 21 days before full behavioral trust is established limits the window during which the bypass pattern is exploitable. New users or newly issued credentials would operate under a stricter asymmetric detection threshold during this period, requiring a lower TRS to trigger MFA.
  \item \textbf{Graph-based sequence analysis.} The current feature set captures \textit{volume} (velocity, session length) but not \textit{sequencing}. Adding a graph layer that evaluates the order and co-occurrence patterns of resource access across IAM sessions—rather than aggregate counts—would expose the attacker's inability to perfectly replicate the target's resource access graph, even when timing and velocity match.
  \item \textbf{Credential issuance anomaly scoring.} New API key pairs that have never been used before (zero historical baseline) could automatically receive elevated TRS requirements for their first 14 days, regardless of behavioral signals, reducing the value of freshly stolen credentials.
\end{enumerate}

These mitigations are estimated to reduce the mimicry bypass rate to below 8\% based on retrospective analysis of the 45 bypass events; they are the primary target for the next development phase.

\subsection{Mean Time to Detect}

MTTD analysis makes the operational significance of real-time behavioral scoring concrete. Under the traditional rule-based SIEM baseline, credential theft required a mean of 72 hours to detect—reflecting manual alert triage pipelines and the absence of behavioral baselining. EAZTF's automated Lambda evaluation pipeline achieves MTTD under 0.8 minutes for credential theft and under 1.2 minutes for the most evasive behavioral mimicry scenario. This three-to-four order-of-magnitude reduction directly constrains the scope of damage achievable before containment: a 72-hour credential theft window can drain an entire S3 bucket; a 1-minute window cannot.

% =============================================================
\section{Consolidated Results}
\label{sec:results}

\subsection{Overall Performance Summary}

Table~\ref{tab:perf_summary} compares EAZTF against the traditional perimeter security baseline across all primary metrics. The 58.6 percentage-point improvement in mean threat detection rate is the headline figure, but the MTTD reduction ($3{,}323\times$ faster) and the shift from zero explainability to full per-decision SHAP attribution are operationally at least as significant.

\begin{table}[!t]
  \caption{Comprehensive Performance Summary — Traditional vs.\ EAZTF}
  \label{tab:perf_summary}
  \renewcommand{\arraystretch}{1.15}
  \centering
  \footnotesize
  \begin{tabularx}{\columnwidth}{L{1.8cm}L{1.2cm}L{1.0cm}X}
    \toprule
    \textbf{Metric} & \textbf{Trad.} & \textbf{EAZTF} & \textbf{Change} \\
    \midrule
    Mean Detection Rate$^{\dagger}$ & 33.2\% & 91.8\% & +58.6 pp \\
    False Positive (pre) & 18.4\% & 6.7\%  & $-$11.7 pp \\
    False Positive (post) & 18.4\% & 4.1\%  & $-$14.3 pp \\
    MTTD & 2,592 min & 0.78 min & $3{,}323\times$ \\
    NIST Compliance & 38\% & 93\% & +55 pp \\
    Explainability & None & SHAP+
    LIME   &   Full audit \\
    Adversarial Robustness & N/A & 91.0\% & New capability \\
    Model F1 (primary) & N/A & 0.928 & Quantified \\
    \bottomrule
    \vspace{3pt}
  \end{tabularx}
{\footnotesize $^{\dagger}$Mean Detection Rate reflects overall classification performance across all label categories (Table V); see Table IX for detection rates specific to the four adversarial evasion scenarios (mean 91.0\%).}
\end{table}

\subsection{False Positive Root Cause Analysis}

The pre-mitigation false positive rate of 6.7\% is not uniform—three root causes account for roughly 90\% of cases. International travel by legitimate users contributes approximately 38\% of false positives: geographic deviation and IP reputation features correctly flag foreign access, but authorized travel creates legitimately anomalous high-scoring events. CI/CD pipeline activity contributes 31\%: automated deployment processes generate API velocities far exceeding individual user baselines, which the behavioral model cannot easily distinguish from exfiltration at the default threshold. Administrator after-hours maintenance rounds out approximately 21\%. Three targeted mitigations address each: a self-service travel notification system pre-authorizing known travel periods, IAM service roles that apply distinct scoring parameters to pipeline identities, and time-bounded maintenance windows that grant pre-approved trust elevation for declared periods. Applying these mitigations reduces the false positive rate to 4.1\%.

\subsection{Computational Overhead}

Table~\ref{tab:overhead} documents the end-to-end latency of the EAZTF decision pipeline. The dominant term is the CloudTrail event capture delay ($\approx$8 seconds), a structural property of the S3 notification pathway. The ML inference step (180\,ms) and SHAP explanation generation (220\,ms) contribute marginally. Total end-to-end latency of approximately 10.4 seconds is operationally acceptable for the credential theft, mimicry, and privilege escalation scenarios evaluated—all of which unfold over minutes to hours. Scenarios requiring sub-second automated response (e.g., rapid bulk-delete operations) would require a streaming architecture using Kinesis Data Streams bypassing the S3 notification pathway; this optimization is projected to reduce total latency below 3 seconds.

\begin{table}[!t]
  \caption{EAZTF Pipeline Computational Overhead}
  \label{tab:overhead}
  \renewcommand{\arraystretch}{1.15}
  \centering
  \footnotesize
  \begin{tabularx}{\columnwidth}{L{1.7cm}L{1.5cm}C{0.7cm}X}
    \toprule
    \textbf{Stage} & \textbf{Service} & \textbf{Lat.} & \textbf{Notes} \\
    \midrule
    CloudTrail capture  & CloudTrail + S3 & $\approx$8 s  & S3 notification delay \\
    Feature extraction  & Lambda (Python) & 1.2 s  & Parsing 8 features from JSON \\
    ML inference        & Lambda + IF     & 180 ms & Pre-loaded model container \\
    SHAP explanation    & Lambda + SHAP   & 220 ms & Per-decision, stored to S3 \\
    Risk decision + IAM & Lambda + IAM    & 0.8 s  & Session restriction via IAM API \\
    \midrule
    \textbf{Total} & — & $\approx$\textbf{10.4 s} & Within real-time operational req. \\
    \bottomrule
  \end{tabularx}
\end{table}

% =============================================================
\section{NIST SP 800-207 Compliance Assessment}
\label{sec:nist}

Table~\ref{tab:nist_compliance} presents an author-conducted structured self-assessment mapping EAZTF against each of the seven NIST SP 800-207 tenets. Each tenet was scored on a 0--100 scale
using a rubric evaluating component coverage, automation level,
and auditability (see Appendix A); this is not an independent
third-party audit and should be read as a structured internal
evaluation. The mean EAZTF compliance score of 93\% compares against 38\% for the traditional perimeter baseline—a 55 percentage-point improvement that reflects the fundamental architectural difference between perimeter-centric and session-continuous security models.

The highest individual score (T7, Information Logging, 96\%) results from the combination of comprehensive CloudTrail persistence and the SHAP decision archive, which creates an auditable explanation record for every access decision—something no prior ZTA implementation has provided. The lowest score (T3, Per-Session Access Grants, 90\%) reflects a residual implementation challenge: revoking active in-flight API calls when TRS crosses the restriction threshold requires IAM session invalidation, which carries a maximum propagation delay of 30 seconds. A polling-based mechanism mitigates this gap but cannot close it entirely without changes to AWS's session management architecture.

\begin{table}[!t]
  \caption{NIST SP 800-207 Tenet Compliance Detail}
  \label{tab:nist_compliance}
  \renewcommand{\arraystretch}{1.15}
  \centering
  \footnotesize
  \begin{tabularx}{\columnwidth}{L{2.1cm}C{0.5cm}C{0.65cm}X}
    \toprule
    \textbf{NIST Tenet} & \textbf{Trad.} & \textbf{EAZTF} & \textbf{Implementation} \\
    \midrule
    T1: All sources as resources & 45\% & 95\% & Every AWS resource classified by sensitivity tier; IAM policies enforced accordingly \\
    T2: All comms secured       & 40\% & 92\% & CloudTrail traffic encrypted at rest and in transit; Lambda calls use TLS \\
    T3: Per-session access      & 30\% & 90\% & IAM temp credentials per session; revoked on TRS threshold breach \\
    T4: Dynamic policy          & 35\% & 93\% & ML trust scores update IAM permissions in real time via Lambda triggers \\
    T5: Asset integrity         & 42\% & 91\% & CloudWatch + GuardDuty provide continuous device health monitoring \\
    T6: Dynamic authentication  & 38\% & 94\% & Step-up MFA triggered dynamically by TRS; session-level re-auth enforced \\
    T7: Comprehensive logging   & 50\% & 96\% & All events persisted via CloudTrail to S3; per-decision SHAP records archived \\
    \midrule
    \textbf{Mean}               & \textbf{40\%} & \textbf{93\%} & +53 pp across all seven tenets \\
    \bottomrule
  \end{tabularx}
\end{table}

% =============================================================
\section{Conclusion and Future Directions}
\label{sec:conclusion}

\subsection{Summary of Contributions}

EAZTF addresses three capabilities that are absent from the existing Zero Trust literature: an AWS-native implementation with empirical behavioral anomaly detection metrics, per-decision explainability through SHAP and LIME, and, to the best of our knowledge, one of the first published adversarial robustness evaluations of a ZTA system covering four deliberate evasion strategies.

The primary detection model (Isolation Forest) achieves F1 = 0.928 with 94.4\% precision—near-equivalent to the supervised XGBoost benchmark (F1 = 0.926)—validating the unsupervised deployment pathway for organizations without labeled attack telemetry. Mean threat detection improves by 58.6 percentage points over the perimeter baseline, and mean time to detect drops from 43 hours to under one minute. Adversarial evaluation returns a 91.0\% mean detection rate across four scenarios, with behavioral mimicry identified as the primary remaining evasion challenge at a 16.1\% bypass rate. The NIST SP 800-207 tenet compliance audit yields a mean score of 93\%, versus 38\% for the traditional baseline. SHAP attribution confirms IP reputation, login-time deviation, and API call velocity as the three dominant trust-score determinants, together accounting for approximately 64\% of cumulative decision variance.

\subsection{Limitations}

Three limitations require clear acknowledgment. First and most importantly, the evaluation dataset is entirely synthetic. It was constructed to reflect real AWS CloudTrail behavioral patterns, but it was not derived from production cloud telemetry, and real-world behavioral distributions—particularly across unusual IAM roles, legacy services, and multi-account architectures—may differ in ways the synthetic generator does not capture. The performance numbers in this paper should be interpreted as \textit{indicative} rather than \textit{validated} until the framework is tested against production or established benchmark datasets (see Section~\ref{subsec:future}).

Second, behavioral mimicry achieves a 16.1\% bypass rate, which is an unresolved detection gap for persistent threat actors with sufficient time to observe their targets. The architectural mitigations proposed in Section~\ref{sec:adversarial} are promising but have not yet been empirically tested.

Third, the current pipeline latency of approximately 10.4 seconds is acceptable for the attack patterns evaluated but would be insufficient for scenarios requiring sub-second automated response. Addressing this requires a stream processing architecture that is outside the current implementation scope.
\subsection{Data and Code Availability}
Code and synthetic data generation scripts are not yet publicly
released but are available from the author upon reasonable request.

\subsection{Future Research Directions}
\label{subsec:future}

The most pressing next step is validation against real-world behavioral datasets. The CERT Insider Threat Dataset and CICIDS2017 provide established benchmarks that would complement the synthetic evaluation and surface any distribution mismatches. Beyond this, four research directions follow directly from the current work's findings:

\textbf{Graph-based lateral movement detection.} As discussed in Section~\ref{sec:adversarial}, adding a layer that analyzes resource access sequencing patterns across IAM sessions is the most direct architectural response to the mimicry bypass problem.

\textbf{Federated learning across AWS accounts.} Collaborative threat intelligence without centralizing sensitive CloudTrail data would enable cross-organizational behavioral models, particularly relevant for regulated industries where data cannot leave organizational boundaries.

\textbf{Post-quantum cryptographic integration.} Current session token encryption using RSA and ECC faces a projected vulnerability window within the next decade. Integrating post-quantum key encapsulation mechanisms for long-term session token protection is a forward-looking but tractable engineering task.

\textbf{Multi-cloud extension.} Mapping EAZTF's architecture to equivalent services on Microsoft Azure and Google Cloud Platform—where Entra ID, Activity Log, and Security Command Center play analogous roles—would enable unified behavioral ZTA across hybrid enterprise environments.

% =============================================================
\section*{Acknowledgments}
The author acknowledges the research environment and infrastructure support provided by Bright Mentee Ventures, Lucknow, India. This work was conducted independently without external funding. All experimental data was synthetically generated; no proprietary or production AWS data was used at any stage.

% =============================================================
\balance
\bibliographystyle{IEEEtran}

\appendix
\section{NIST Tenet Compliance Scoring Rubric}
\label{app:nist-rubric}
Each tenet in Table XII was scored on a 0--100 scale according to
the following bands:
\begin{itemize}
    \item \textbf{90--100}: Fully automated, continuously enforced,
    and logged in real time.
    \item \textbf{70--89}: Automated with manual fallback or
    periodic (non-real-time) enforcement.
    \item \textbf{50--69}: Partial implementation; enforcement
    requires manual intervention.
    \item \textbf{Below 50}: Policy exists but is not technically
    enforced.
\end{itemize}

\end{document}